\documentclass[aps,pra,10pt,twocolumn,superscriptaddress]{revtex4-1}
\usepackage{amsmath}
\usepackage{amsfonts}
\usepackage{amssymb}
\usepackage{dsfont}
\usepackage{graphicx}
\usepackage{color}
\usepackage{hyperref}
\hypersetup{colorlinks=true,allcolors=blue}

\newcommand{\bq}{\mathbf{q}}
\newcommand{\st}[1]{_\text{#1}} 
\newcommand{\BRA}[1]{\langle #1\vert} 
\newcommand{\KET}[1]{\vert #1\rangle} 
\newcommand{\OP}[2]{\KET{#1}\BRA{#2}} 
\newcommand{\PRO}[1]{\OP{#1}{#1}} 
\newcommand{\ABS}[1]{\vert #1\vert} 
\newcommand{\ddt}{\frac{\mathrm{d}}{\mathrm{d}t}} 
\newcommand{\LB}[2]{\mathcal{L}_{#1,#2}} 
\newcommand{\Tr}[1]{\mathrm{Tr}\left\lbrace #1 \right\rbrace} 

\newcommand{\dcL}[1]{\Delta\st{#1}}

\begin{document}

\title{Phonon-induced transition between entangled and nonentangled photon emission in constantly driven quantum-dot--cavity systems}

\author{T. Seidelmann}
\email[Corresponding author: ]{tim.seidelmann@uni-bayreuth.de}
\affiliation{Lehrstuhl f{\"u}r Theoretische Physik III, Universit{\"a}t Bayreuth, 95440 Bayreuth, Germany}
\author{M. Cosacchi}
\affiliation{Lehrstuhl f{\"u}r Theoretische Physik III, Universit{\"a}t Bayreuth, 95440 Bayreuth, Germany}
\author{M. Cygorek}
\affiliation{Heriot-Watt University, Edinburgh EH14 4AS, United Kingdom}
\author{D. E. Reiter}
\altaffiliation[Current address: ]{Condensed Matter Theory, Department of Physics, TU Dortmund, 44221 Dortmund, Germany}
\affiliation{Institut f{\"u}r Festk{\"o}rpertheorie, Universit{\"a}t M{\"u}nster, 48149 M{\"u}nster, Germany}
\author{A. Vagov}
\affiliation{Lehrstuhl f{\"u}r Theoretische Physik III, Universit{\"a}t Bayreuth, 95440 Bayreuth, Germany}
\author{V. M. Axt}
\affiliation{Lehrstuhl f{\"u}r Theoretische Physik III, Universit{\"a}t Bayreuth, 95440 Bayreuth, Germany}

\begin{abstract}
Entangled photon pairs are essential for many applications in quantum technologies. Recent theoretical studies demonstrated that different types of entangled Bell states can be created in a constantly driven four-level quantum emitter-cavity system. Unlike other candidates for the realization of the four-level emitter, semiconductor quantum dots unavoidably interact with their environment, resulting in carrier-phonon interactions.
Surprisingly, phonons change the entanglement of emitted photon pairs in a qualitative way, already at low temperatures on the order of 4~K. While one type of Bell state can still be generated using small driving strengths, the other type is suppressed due to phonon interactions in strongly-confined quantum dots. The degree of entanglement decreases with rising temperature and driving strength until it vanishes at a certain parameter value. Because it remains zero afterward, we encounter a phonon-induced transition between entangled and nonentangled photon emission that resembles a phase transition. The transition occurs at temperatures below 30\,K and, independent of the driving strength, the concurrence as a function of the reduced temperature is found to obey a power law with exponent one near the transition point. 
\end{abstract}

\maketitle

\section{Introduction}
\label{sec:introduction}

The phenomenon of quantum entanglement is one of the most fascinating and unintuitive effects in nature. Being a pure quantum effect, entanglement is interesting not only from a fundamental view, but it also prompted the development of innovative applications in novel research fields, like quantum cryptography \cite{Gisin:02,Lo_quantum_cryptography}, quantum communication \cite{duan_quantum_comm,Huber_overview_2018}, and quantum information processing and computing \cite{pan:12,Bennett:00,Kuhn:16,Zeilinger_entangled}.

An often discussed realization of entangled qubits are polarization entangled photon pairs. Typically, one aims for the generation of one of the four Bell states
\begin{subequations}
\begin{equation}
\label{eq:def_phi_bell_state}
\KET{\Phi_\pm} = \frac{1}{\sqrt{2}}\left( \KET{HH} \pm \KET{VV} \right) ,
\end{equation}
\begin{equation}
\label{eq:def_psi_bell_state}
\KET{\Psi_\pm} = \frac{1}{\sqrt{2}}\left( \KET{HV} \pm \KET{VH} \right) ,
\end{equation}
\end{subequations}
the most prominent maximally entangled states established for linearly polarized photon pairs. Here $H$ and $V$ denote horizontally and vertically polarized photons, respectively, and their order reflects the order of photon detection. One can distinguish between two different types of Bell states (or Bell state entanglement); while in a $\Phi$ Bell state ($\Phi$BS), the first and second detected photon possess the same polarization, in a $\Psi$ Bell state ($\Psi$BS), they exhibit the opposite one. 

A promising platform for the generation of a maximally entangled Bell state are semiconductor quantum dots (QDs), which realize a four-level quantum emitter \cite{Orieux_entangled,Seidelmann_QUTE_2020}.
The biexciton-exciton cascade in QDs comprises its ground state, two exciton states, and the biexciton. Due to the optical selection rules, these four electronic levels display a diamond-shaped configuration, cf., Fig.~\ref{fig:system}. After an initial excitation of the biexciton \cite{entangled-photon1,Finley_phonon-assisted,Reiter_2014,PI_phonon-assisted_biexc_prep-exp,hanschke2018,huber2017,reindl2017}, the subsequent photon emission induced by the cascade, should, in an ideal situation, result in the generation of a maximally entangled $\Phi$BS. By embedding the QD inside a microcavity, one can enhance the light-collection efficiency and photon emission rate due to the Purcell effect \cite{dousse:10}. Furthermore, the energy of the cavity modes can have a profound impact on the resulting type and degree of entanglement \cite{Jahnke2012,delValle_twoPhoton,munoz15,Seidelmann2019,Seidelmann_QUTE_2020}.

Indeed, various theoretical and experimental studies demonstrated the possibility to obtain $\Phi$BS entanglement in the chosen basis of linearly polarized photons \cite{Seidelmann2019,Different-Concurrences:18, Phon_enhanced_entanglement,Jahnke2012, heinze17,BiexcCasc_Carmele,Stevenson2006,Young_2006,Muller_2009,Huber_PRL_2018, Wang_2019,Liu2019, Bounouar18,dousse:10,winik:2017, entangled-photon1,Fognini_2019, entangled-photon2,Hafenbrak, Biexc_FSS_electrical_control_Bennett, EdV,Troiani2006,stevenson:2012,Benson_2000_QD_cav_device,Basset_quantum_key_2021,
Christian_Schimpf_Crypto_2021,Christian_Schimpf_strongly_ent_2021}. Furthermore, recent theoretical studies \cite{munoz15,Seidelmann_QUTE_2020,Seidelmann_Switching} showed that a four-level emitter-cavity system, e.g., a QD embedded inside a cavity, can also facilitate the creation of $\Psi$BS entanglement, when a constant laser driving is applied to the emitter.
In this setup, four laser-dressed states emerge and their characteristics depend on the applied driving strength. By adjusting the cavity modes to a direct two-photon transition between these dressed states, entangled photon pairs can be created.
The resulting type and degree of entanglement depends crucially on the applied driving strength and the energy of the cavity modes \cite{Seidelmann_QUTE_2020,Seidelmann_Switching}.

In contrast to other possible realizations of the quantum emitter, e.g., atomic systems, QDs unavoidably interact with their semiconductor environment, which results in carrier-phonon interactions \cite{besombes:01,pawel04,kaer10,Hughes_theory_spectren,NazirReview,Carmele_2019_NonMarkov,
McCutcheon_nonMarkov,Uebersichtsartikel_2019}. Although phonons are associated with the loss of quantum coherence in the system, their possible impact on the resulting two-photon state and its degree of entanglement cannot be easily predicted. Indeed, various scenarios and dependencies have been found for different QD setups. Although the interaction with phonons often results in a reduced degree of entanglement, their detrimental impact can depend strongly on the chosen cavity arrangement \cite{Seidelmann2019}. Furthermore, in highly symmetric situations, phonons may have no impact on entanglement at all \cite{BiexcCasc_Carmele}. Moreover, even a phonon-induced enhancement of the degree of entanglement has been predicted under special conditions \cite{Phon_enhanced_entanglement}. Therefore, it is not a priori clear in which manner phonons may impact the constantly driven QD-cavity system.

In this paper, we investigate the phonon influence on emitted photon pairs in this system. In our theoretical study, we consider the configuration where the highest degrees of $\Phi$BS and $\Psi$BS entanglement have been predicted in the phonon-free analysis of a four-level quantum emitter \cite{Seidelmann_QUTE_2020}. Remarkably, the phonon impact on the entanglement is much more severe than in studies employing an initially prepared biexciton without constant driving. Not only does the interaction with phonons in strongly-confined QDs reduce the degree of entanglement for small driving strength values already at 4\,K, but it results in the absence of entanglement for higher driving strength values and/or temperatures.
The observed change of the emitted photon states in the presence of phonon interactions can be understood as a competition between two relaxation processes that result in different stationary states related to the emission of either entangled or nonentangled photons.
A qualitative change of the stationary state of a driven dissipative system upon change of a parameter is often referred to as a dissipative phase transition \cite{Minganti_2021,Kirton_2018}.
Here, the encountered transition between entangled and nonentangled photon emission with increasing temperature or driving strength is akin to a phase transition, showing some features typically associated with the latter.

\section{Driven quantum-dot--cavity system}
\label{sec:system}

\begin{figure}[t]
\centering
\includegraphics[width=\columnwidth]{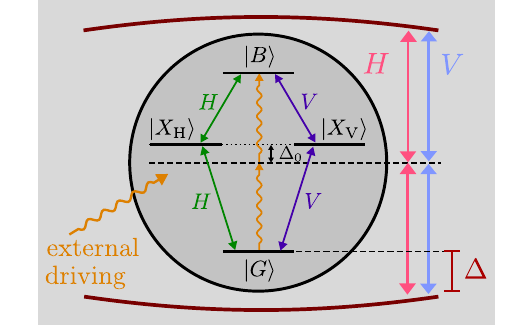}
\caption{
Schematic sketch of the driven QD-cavity system. The biexciton-exciton cascade comprises the ground state $\KET{G}$, two energetically degenerate exciton states $\KET{X\st{H/V}}$ that couple to horizontally or vertically polarized light, respectively (indicated by green and purple arrows), and the biexciton state $\KET{B}$. Curvy orange arrows represent the external laser driving adjusted to the two-photon transition between $\KET{G}$ and $\KET{B}$. The laser is detuned from the exciton energy by the value $\Delta_0=E\st{B}/2$. The energy of the two degenerate but orthogonally polarized cavity modes (red and blue arrows) is described by the cavity-laser detuning $\Delta$.
}
\label{fig:system}
\end{figure}

We consider the biexciton-exciton cascade of a strongly-confined, self-assembled GaAs semiconductor QD. The QD is embedded inside a microcavity and is continuously driven by an external laser, cf., Fig.~\ref{fig:system} for a schematic sketch. Furthermore, the QD interacts with its surrounding semiconductor environment, resulting in a coupling to lattice oscillations, i.e., phonons. In the case of strongly-confined GaAs QDs at low temperatures, the most important interaction is due to the deformation potential coupling to longitudinal acoustic (LA) phonons \cite{dotspektren:02}.

The QD comprises the ground state $\KET{G}$, two orthogonally polarized exciton states $\KET{X\st{H/V}}$ at energy $\hbar\omega\st{X}$ that couple to horizontally ($H$) and vertically ($V$) polarized light, respectively, and the biexciton state $\KET{B}$. In the frame co-rotating with the external laser frequency $\omega\st{L}$, the QD-cavity Hamiltonian is given by \cite{munoz15,Seidelmann_QUTE_2020}
\begin{equation}
\label{eq:H_QD-cav}
\begin{split}
\hat{H}\st{QD-C} &= \Delta_0 \left( \PRO{X\st{H}} + \PRO{X\st{V}} \right) \\
&+ \left( 2\Delta_0-E\st{B} \right) \PRO{B} \\ 
&+ \sum\limits_{\ell = H,V} \Delta \hat{a}_\ell^\dagger\hat{a}_\ell + \sum\limits_{\ell = H,V} g\left(\hat{a}_\ell^\dagger\hat{\sigma}_\ell + \hat{a}_\ell\hat{\sigma}_\ell^\dagger \right) ,
\end{split}
\end{equation}
where the energy of the ground state is used as the zero of the energy scale, $\Delta_0 := \hbar\left(\omega\st{X}-\omega\st{L}\right)$  is the energetic detuning between the exciton states and the laser energy, and $E\st{B}$ denotes the biexciton binding energy. The electronic transitions of the QD are described by the operators
\begin{subequations}
\label{eq:sigmas}
\begin{equation}
\hat{\sigma}\st{H} = \OP{G}{X\st{H}} + \OP{X\st{H}}{B} ,
\end{equation}
\begin{equation}
\hat{\sigma}\st{V} = \OP{G}{X\st{V}} + \OP{X\st{V}}{B}
\end{equation}
\end{subequations}
and are coupled to two energetically degenerate, but orthogonally polarized cavity modes with energy $\hbar\omega\st{C}$. The bosonic operator $\hat{a}\st{H/V}^\dagger$ creates a cavity photon with the respective polarization, $H$ or $V$, and the QD-cavity coupling strength $g$ is assumed to be equal for all transitions. The energy of the cavity modes is described by the cavity-laser detuning $\Delta := \hbar\left(\omega\st{C}-\omega\st{L}\right)$.

An external laser with driving strength $\Omega$ and constant frequency $\omega\st{L}$ continuously excites the QD. Following Ref.~\onlinecite{Seidelmann_QUTE_2020} the frequency is adjusted to the two-photon transition between ground and biexciton state, i.e., $\Delta_0 = E\st{B}/2$, and the polarization is chosen to be diagonal in the basis spanned by $H$ and $V$. In the rotating frame, the respective Hamiltonian is given by
\begin{equation}
\label{eq:H_L}
\hat{H}\st{L} = \Omega \left( \hat{\sigma}\st{D} + \hat{\sigma}\st{D}^\dagger \right);\hspace{.5cm} \hat{\sigma}\st{D} = \left( \hat{\sigma}\st{H}+\hat{\sigma}\st{V} \right)/\sqrt{2}.
\end{equation}

The coupling to LA phonons is described by
\begin{equation}
\label{eq:H_Ph}
\hat{H}\st{Ph} = \hbar\sum\limits_\bq \omega_\bq\hat{b}_\bq^\dagger\hat{b}_\bq + \hbar\sum\limits_{\chi,\bq} n_\chi \left( \gamma_\bq^X\hat{b}_\bq^\dagger + {\gamma_\bq^X}^\ast\hat{b}_\bq \right) \PRO{\chi} ,
\end{equation}
where the bosonic operator $\hat{b}_\bq$ destroys a phonon in mode $\bq$ with energy $\hbar\omega_\bq$. $\gamma_\bq^X$ denotes the exciton-phonon coupling strength and $n_\chi=\lbrace 0,1,1,2 \rbrace$ is the number of excitons present in the QD state $\KET{\chi}\in\left\lbrace\KET{G},\KET{X\st{H}},\KET{X\st{V}},\KET{B}\right\rbrace$.

Furthermore, important loss channels, namely cavity losses with rate $\kappa$ and radiative decay with rate $\gamma$, are incorporated into the model via Lindblad operators \cite{Lindblad:1976}
\begin{equation}
\label{eq:Lindblad}
\LB{\hat{O}}{\Gamma}\hat{\rho} = \frac{\Gamma}{2}\left(2\hat{O}\hat{\rho}\hat{O}^\dagger - \hat{O}^\dagger\hat{O}\hat{\rho} - \hat{\rho}\hat{O}^\dagger\hat{O} \right) ,
\end{equation}
where $\hat{O}$ is the system operator associated with a loss process with rate $\Gamma$.

The dynamics of the statistical operator of the system $\hat{\rho}$ is described by the Liouville-von~Neumann equation
\begin{eqnarray}
\label{eq:LvN_Eq}
\ddt\hat{\rho} &=& \mathcal{L}\hat{\rho} :=-\frac{i}{\hbar}\left[\hat{H},\hat{\rho}\right] \\
&&+ \sum\limits_{\ell=H,V} \left\lbrace \LB{\hat{a}_\ell}{\kappa} + \LB{\OP{G}{X_\ell}}{\gamma} + \LB{\OP{X_\ell}{B}}{\gamma} \right\rbrace \hat{\rho} \notag \\
\label{eq:Hamiltonian}
\hat{H} &=& \hat{H}\st{QD-C} + \hat{H}\st{L} + \hat{H}\st{Ph} ,
\end{eqnarray}
where $[\cdot,\cdot]$ denotes the commutator. Employing a real-time path-integral method (consult Refs.~\onlinecite{Makri_Theory,Makri_Numerics,PI_realtime2011,PI_nonHamil2016,PI_cQED} for details) the time-evolution of the reduced density matrix of the QD-cavity system is evaluated in a numerically exact manner. For our numerical calculations, we assume that the phonons are initially in thermal equilibrium at temperature $T$ and that the QD-cavity system is initially in the ground state $\KET{G}$ without any cavity photons.

Following Refs.~\onlinecite{munoz15,Seidelmann_QUTE_2020} we choose realistic parameters for the QD-cavity system that are summarized in Table~\ref{tab:Fixed_Parameters}. Furthermore, we consider a spherically symmetric GaAs QD with a harmonic oscillator confinement and an electron (hole) confinement length $a\st{e}=3$~nm ($a\st{h} = a\st{e}/1.15$). The deformation potential coupling of the QD to LA phonons enters the path-integral calculations via the phonon spectral density $J(\omega) = \sum_\bq \ABS{\gamma_\bq^X}^2\,\delta(\omega-\omega_\bq)$. An explicit expression for this quantity, assuming a linear dispersion relation, and the used material parameters can be found in Appendix~\ref{app:gaas_density}.

\begin{table}
\centering
\caption{Fixed system parameters used in the calculations.}
\label{tab:Fixed_Parameters}
\begin{ruledtabular}
\begin{tabular}{l c c}
Parameter & & Value\\
\hline
QD-cavity coupling strength & $g$ & $0.051$~meV \\
Biexciton binding energy & $E\st{B}$ & $20g=1.02$~meV \\
Detuning & $\Delta_0$ & $E\st{B}/2=0.51$~meV \\
Cavity loss rate & $\kappa$ & $0.1g/\hbar \approx 7.8$~$\mathrm{ns^{-1}}$\\
Radiative decay rate & $\gamma$ & $0.01g/\hbar \approx 0.78$~$\mathrm{ns^{-1}}$\\
\end{tabular}
\end{ruledtabular}
\end{table}

\section{Entanglement determination}
\label{sec:entanglement}

\subsection{Two-photon density matrix}
\label{subsec:TPDM}

In a typical experimental setup the two-photon density matrix $\rho^\text{2p}$ is reconstructed using quantum state tomography \cite{QuantumStateTomography}. This reconstruction scheme relies on polarization-resolved two-time correlation measurements. The detected signals in these measurements are proportional to two-time correlation functions
\begin{equation}
\label{eq:G2}
G_{jk,\ell m}^{(2)}(t,\tau^\prime) = \left\langle\hat{a}_j^\dagger(t)\hat{a}_k^\dagger(t+\tau^\prime)\hat{a}_m(t+\tau^\prime)\hat{a}_\ell(t)\right\rangle ,
\end{equation}
where $\left\lbrace j,k,\ell,m \right\rbrace\in\left\lbrace H,V \right\rbrace$. Here, $t$ is the time when the first photon is detected and $\tau^\prime$ the delay time until a subsequent, second photon is detected. Although Eq.~\eqref{eq:G2} describes cavity photons, it can also be used to model correlation functions for photons measured in the free space outside the cavity, when the outcoupling of light from the cavity into the free space is assumed to be a Markovian process \cite{Kuhn:16}. For details on the evaluation of multi-time correlation functions in the path-integral framework, we refer to Ref.~\onlinecite{multi-time}.

In standard experiments the measurement data is typically averaged over finite real and delay time intervals. Thus, the reconstructed two-photon density matrix is theoretically calculated as \cite{munoz15,Seidelmann_QUTE_2020}
\begin{subequations}
\label{eq:2pht_dm}
\begin{equation}
\rho_{jk,\ell m}^\text{2p} = \frac{\overline{G}_{jk,\ell m}^{(2)}}{\Tr{\overline{G}^{(2)}}} ,
\end{equation}
\begin{equation}
\overline{G}_{jk,\ell m}^{(2)} = \frac{1}{\Delta t\,\tau} \int\limits_{t_0}^{t_0+\Delta t}\mathrm{d}t \int\limits_0^\tau \mathrm{d}\tau^\prime\,G_{jk,\ell m}^{(2)}(t,\tau^\prime) ,
\end{equation}
\end{subequations}
where $t_0$ is the starting time of the coincidence measurement and $\tau$ ($\Delta t$) the used delay time (real time) window. The trace $\Tr{\cdot}$ in Eq.~\eqref{eq:2pht_dm} is introduced for the purpose of normalization. Note that, in principle, $\rho^\text{2p}$ depends on all three measurement parameters: $t_0$, $\Delta t$, and $\tau$ \cite{Different-Concurrences:18}.

Throughout this paper, the two-photon density matrix is determined for the steady state $\hat{\rho}_s$ of the system defined as $\ddt\hat{\rho}_s=\mathcal{L}\hat{\rho}_s=0$. Thus, in the calculation scheme, the time $t_0$ is chosen such that it occurs after the system has reached a stationary density matrix in the time evolution. Note that in this situation $\overline{G}_{jk,\ell m}^{(2)}$ and therefore also $\rho^\text{2p}$ become independent of $t_0$ and $\Delta t$. Thus the two-photon density matrix does only depend on the delay time window $\tau$. In general, different delay time windows correspond to selecting different two-photon subsets in the total emission from the QD-cavity system \cite{Seidelmann2019}. Although  the photon yield decreases with shorter windows $\tau$, the degree of entanglement typically increases in this situation \cite{munoz15,Seidelmann2019}. In our study we thus fix the delay time window to a short, but realistic value of $\tau=50$~ps, which we adopt from Ref.~\onlinecite{StevensonPRL2008}.

Altogether, the described reconstruction and calculation scheme gives the two-photon density matrix $\rho^\text{2p}$ in the basis $\left\lbrace \KET{HH},\KET{HV},\KET{VH},\KET{VV}\right\rbrace$.
Besides the polarization degree of freedom, the two recorded photons in these states also differ in their emission and detection times.
This means, while in the two-photon state $\KET{HV}$, the first detected photon is $H$ polarized and the subsequent second one is $V$ polarized, this order is exactly reversed for $\KET{VH}$. Note that, since a finite delay time window $\tau$ is considered, the situation where both photons are detected simultaneously (i.e., $\tau^\prime=0$) is a subset of measure zero and the two-photon states $\KET{HV}$ and $\KET{VH}$ can indeed be distinguished, for all relevant delays $\tau'\neq 0$.

After the two-photon density matrix has been obtained as described above, the type of entanglement can be determined directly from its form. One encounters a $\Phi$BS ($\Psi$BS) when the corresponding occupations of the two-photon states $\KET{HH}$ and $\KET{VV}$ ($\KET{HV}$ and $\KET{VH}$) dominate. 

\subsection{Concurrence}
\label{subsec:concurrence}

The degree of entanglement associated with a given two-photon density matrix $\rho^\text{2p}$ is quantified by the concurrence \cite{Wootters1998}. The concurrence $C$ has a one-to-one correspondence to the entanglement of formation, which in turn represents the minimal amount of pure-state entanglement that is at least present in a mixed state described by a given two-qubit density matrix \cite{Wootters1998,Different-Concurrences:18}. In contrast to the latter, the concurrence can be calculated directly from the two-photon density matrix $\rho^\text{2p}$ according to \cite{QuantumStateTomography,Seidelmann_QUTE_2020,Wootters1998}
\begin{equation}
\label{eq:Concurrence}
C = \max\left\lbrace 0,\sqrt{\lambda_1}-\sqrt{\lambda_2}-\sqrt{\lambda_3}-\sqrt{\lambda_4}\right\rbrace ,
\end{equation}
where $\lambda_j \geq \lambda_{j+1}$ are the (real and positive) eigenvalues of the matrix
\begin{equation}
\label{eq_matrix_M}
M = \rho^\text{2p}\,T\,(\rho^\text{2p})^\ast\,T .
\end{equation}
Here, $T$ is an anti-diagonal 4$\times$4-matrix with elements $\left\lbrace -1,1,1,-1\right\rbrace$ and $(\rho^\text{2p})^\ast$ denotes the Hermitian conjugated two-photon density matrix. Because $\rho^\text{2p}$ depends in principle on the parameters $t_0$, $\Delta t$, and $\tau$, the same applies to the concurrence.

Altogether, the type and degree of entanglement is obtained as follows: (i) The averaged two-photon correlations $\overline{G}_{jk,\ell m}^{(2)}$ are calculated for the steady state employing the path-integral method. (ii) The two-photon density matrix $\rho^\text{2p}$ is calculated according to Eq.~\eqref{eq:2pht_dm} and the type of entanglement can be identified. (iii) The concurrence is evaluated for the obtained density matrix $\rho^\text{2p}$ using Eq.~\eqref{eq:Concurrence}.

\section{Result: Phonon-induced transition}
\label{sec:phase_transition}

\subsection{Phonon-free results}

Because the parameter space of the considered system is quite large, we restrict our study to the parameters, where the highest degrees of entanglement are obtained in the phonon-free case. In Ref.~\onlinecite{Seidelmann_QUTE_2020}, the phonon-free case was analyzed in detail for a general four-level quantum emitter and it was demonstrated that the resulting type of entanglement and its degree depend on the applied driving strength $\Omega$ and the used cavity-laser detuning $\Delta$.

\begin{figure}[t]
\centering
\includegraphics[width=\columnwidth]{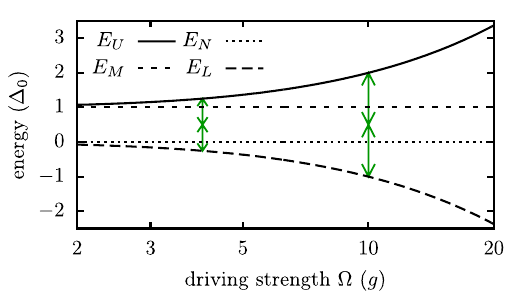}
\caption{
Energies of the four laser-dressed states $\KET{U}$, $\KET{M}$, $\KET{N}$, and $\KET{L}$ as a function of the driving strength $\Omega$. The cavity modes are always tuned to match the two-photon resonance between $\KET{U}$ and $\KET{L}$. For two exemplary driving strength values, the energy of the cavity modes $\Delta$ (in a frame co-rotating with the laser frequency) is indicated as green double-headed arrows.
}
\label{fig:dressed_states}
\end{figure}

Due to the constant laser excitation, transitions do not take place between the bare states $\KET{G}$, $\KET{X\st{H/V}}$, and $\KET{B}$ but rather between  eigenstates of the constantly driven QD, i.e., the four laser-dressed states which we denote as $\KET{U}$ (``uppermost''), $\KET{M}$ (``middle''), $\KET{N}$ (``null''), and $\KET{L}$ (``lowest''). The corresponding eigenenergies are given by \cite{Seidelmann_QUTE_2020}
\begin{subequations}
\label{eq:laser_dressed energies}
\begin{eqnarray}
E\st{U} &=& \frac{1}{2} \left( \Delta_0 + \sqrt{\Delta_0^2 + 8\Omega^2} \right) \\
E\st{M} &=& \Delta_0 \\
E\st{N} &=& 0 \\
E\st{L} &=& \frac{1}{2} \left( \Delta_0 - \sqrt{\Delta_0^2 + 8\Omega^2} \right)
\end{eqnarray}
\end{subequations}
and depicted in Fig.~\ref{fig:dressed_states}. Note that these energies, and in turn the transition energies between them, depend on the driving strength.

The detailed analysis of the phonon-free situation in Ref.~\onlinecite{Seidelmann_QUTE_2020} showed that the type of entanglement - $\Phi$BS or $\Psi$BS - and its degree depend on the driving strength and the cavity mode placement. Essentially, a high degree of entanglement is only possible if the cavity modes are close to resonance with a direct two-photon transition between two of the four laser-dressed states, i.e., $\Delta = (E_\chi-E_{\chi^\prime})/2$ where $\chi\neq\chi^\prime$ denote any pair of laser-dressed states. Thus several possibilities for the cavity-laser detuning $\Delta$ were analyzed. It was found that, while the type of entanglement remains the same for some of the two-photon resonances, it can change with the driving strength for others. However, because neighboring two-photon resonances can affect each other, the highest degrees of entanglement, especially in the case of $\Psi$BS entanglement, were obtained for the two-photon resonance between the uppermost and lowest laser-dressed states, as this resonance condition is typically quite separated from the others.

Thus, in this study, this resonance condition is chosen for further investigation. It is selected by adjusting the cavity-laser detuning to \cite{Seidelmann_QUTE_2020}
\begin{equation}
\label{eq:fix_Delta}
\Delta = \frac{\Delta\st{UL}}{2} := \frac{E\st{U}-E\st{L}}{2} = \frac{1}{2}\sqrt{\Delta_0^2+8\Omega^2} ,
\end{equation}
which corresponds to tuning the cavity mode energy to $ \hbar\omega_C = \Delta\st{UL}/2-\Delta_0+\hbar\omega_X $. Note that the cavity-laser detuning in Eq.~\eqref{eq:fix_Delta} depends on the external driving strength $\Omega$. Thus, in the following, this detuning is changed alongside $\Omega$ in order to keep the desired two-photon resonance condition, cf., Fig.~\ref{fig:dressed_states}.

\begin{figure*}[t]
\centering
\includegraphics[width=\textwidth]{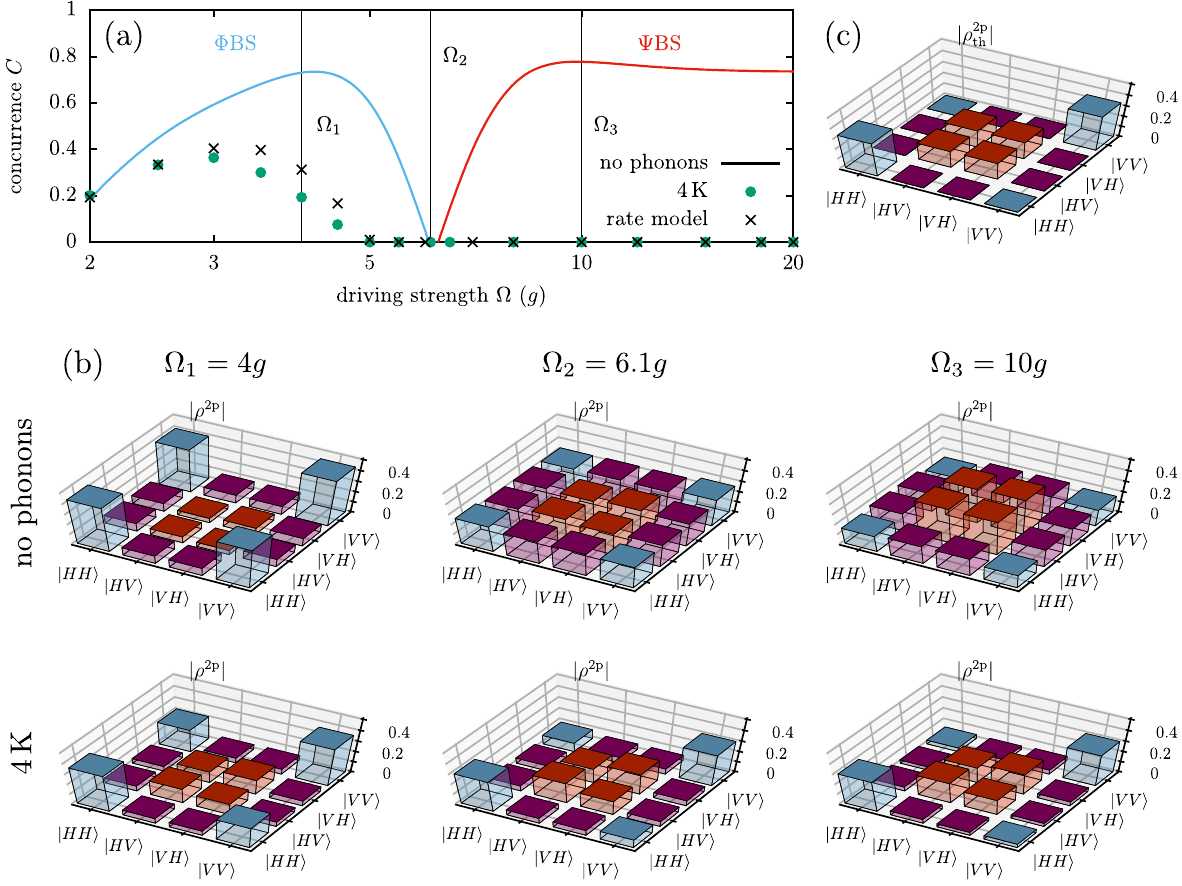}
\caption{
(a) Concurrence as a function of the driving strength $\Omega$ (in units of the coupling strength $g$) without phonons (solid line) and with phonons at a temperature of $4\,$K (green dots). The phonon-free curve is color coded: blue (red) indicates $\Phi$BS ($\Psi$BS) entanglement. Three driving strength values, that are associated with either a high concurrence or a vanishing degree of entanglement in the phonon-free situation, are marked with straight vertical lines. Note that the cavity-laser detuning $\Delta = \Delta\st{UL}/2 =\sqrt{\Delta_0^2+8\Omega^2}/2$ is changed alongside $\Omega$ to keep the desired resonance condition. Black crosses depict the results obtained with a rate model according to Eq.~\eqref{eq:rate_model} based on the rates given in Fig.~\ref{fig:rates}. 
(b) Corresponding two-photon density matrices (absolute values) for the three driving strengths $\Omega_j$ indicated in panel (a). Results are shown for calculations without phonons (upper row) and including phonons at $T=4\,$K (lower row).
(c) Two-photon density matrix calculated from the thermally distributed state as described in Eqs.~\eqref{eq:rho_therm_UL} and \eqref{eq:2pdm_therm_UL}. 
}
\label{fig:phase_transition}
\end{figure*}

Figure~\ref{fig:phase_transition}(a) depicts the concurrence as a function of the driving strength $\Omega$ at the two-photon transition between the laser-dressed states $\KET{U}$ and $\KET{L}$. The results in the phonon-free situation (solid line) are in accordance with those presented in Ref.~\onlinecite{Seidelmann_QUTE_2020}. A region of high $\Phi$BS ($\Psi$BS) entanglement, indicated by blue (red) curve segments, is found for low (high) driving strength values. In between the two regions of high entanglement, a special point occurs at $\Omega\st{sp}=\sqrt{3/8}\Delta_0\approx 6.12 g$ where the concurrence drops to zero. The corresponding two-photon density matrix, calculated for three selected driving strength values $\Omega_j$ and illustrated in the upper row of Fig.~\ref{fig:phase_transition}(b), clearly shows the transition from a state that is close to a maximally entangled $\Phi$BS (cf., $\Omega_1$) to an entangled $\Psi$BS with high concurrence (cf., $\Omega_3$).

The reason for this behavior is analyzed in detail in Ref.~\onlinecite{Seidelmann_QUTE_2020}. With increasing driving strength the composition of the individual laser-dressed states changes, resulting in changing optical selection rules between them. Around the driving strength $\Omega_1$, the system emits predominantly two equally polarized photons when a direct transition between the uppermost dressed state into the lowest one occurs and, consequently, one finds a $\Phi$BS. With rising driving strength the probability for the simultaneous emission of two photons with opposite polarizations increases. At the driving strength $\Omega\st{sp}=\sqrt{3/8}\Delta_0$, both type of processes have the same probability and the degree of entanglement vanishes. When the driving strength is increased beyond this point, the latter becomes more and more dominant until a direct transition between the dressed states $\KET{U}$ and $\KET{L}$ is almost exclusively accompanied by the emission of two photons with opposite polarization. Thus one obtains a $\Psi$BS in this regime.

\subsection{Results including LA phonons}

This well understood behavior changes drastically when the interaction to LA phonons is included in the calculations [green dots in Fig.~\ref{fig:phase_transition}(a)], already at a very low temperature of $T=4$\,K. Besides a reduced concurrence at small driving strength values and the fact that a maximum concurrence of only $C\approx 0.4$ is reached, the interaction with LA phonons also alters the curve qualitatively.

Although, for small $\Omega$, the concurrence follows the phonon-free result, it starts to decrease already after reaching a maximum around $\Omega\approx 3g$, and drops to zero well before the special point at $\Omega\approx\Omega_2$. In stark contrast to the phonon-free results, the concurrence remains zero for $\Omega > \Omega_2$ and no subsequent region of neither $\Phi$BS nor $\Psi$BS entanglement emerges. It is important to stress that LA phonons do not fully destroy the degree of entanglement in systems without constant laser driving, especially when the fine-structure splitting between the exciton states is zero \cite{Seidelmann2019}. In this special situation, theoretical calculations of the dynamics of initially prepared biexciton states even predict a maximally entangled two-photon state, even when phonons are taken into account \cite{BiexcCasc_Carmele,Seidelmann2019}.

In contrast, the constantly driven system demonstrates a behavior similar to a phase transition, where an entangled photon pair is generated only when the driving strength is below a certain value. Above this value the degree of entanglement, as measured by the concurrence, remains strictly zero.
In the following we provide an explanation for the observed transition between entangled and nonentangled photon emission. To this end, we first examine the two-photon density matrix at $T=4$\,K and infer the resulting concurrence. In a second step, we consider the role of LA phonons and explain why they lead to the obtained form of the density matrix. Finally, we turn to the impact of rising temperature, where the behavior of the concurrence again resembles a phase transition. 

\subsection{Concurrence and density matrices at 4~K}
\label{subsec:2pdm_at_4K}

The second row of Fig.~\ref{fig:phase_transition}(b) depicts the two-photon density matrix at $T=4$\,K obtained for three selected driving strength values $\Omega_j$. In the case of the small driving strength $\Omega_1=4g$, the characteristics of $\rho^\text{2p}$ resemble the two-photon density matrix in the phonon-free case. The occupations of the states $\KET{HH}$ and $\KET{VV}$ and their coherences (blue bars) dominate over the remaining elements. Thus one detects a finite degree of $\Phi$BS entanglement. A more detailed comparison to the phonon-free result reveals that the coherence associated with the matrix element $\OP{HH}{VV}$ is reduced by a factor of 2 due to the phonon interaction, and the occupations and coherences associated with a $\Psi$BS (red bars) are enhanced at the expense of the blue bars. These two effects combined lead to a finite, but reduced degree of entanglement. When the driving strength is increased at $T=4$\,K, similar trends can be observed if one compares the subsequent density matrices for $\Omega_1$, $\Omega_2$, and $\Omega_3$. With increasing driving strength, the coherence $\OP{HH}{VV}$ is further reduced, while the elements linked to a $\Psi$BS (red bars) increase.

Disregarding other (insignificant) coherences (purple bars), the two-photon density matrices at $T=4$\,K can all be well represented by matrices with 4 independent entries having the form
\begin{equation}
\label{eq:special_form}
\rho^\text{2p} = \begin{pmatrix}
a & 0 & 0 & c\\
0 & b & d & 0\\
0 & d^\ast & b & 0\\
c^\ast & 0 & 0 & a
\end{pmatrix} ,
\end{equation}
where the parameters fulfill the requirements for an arbitrary density matrix
\begin{equation}
a,b\in\mathbb{R}^+_0;\hspace{.15cm}2(a+b)=1;\hspace{.15cm}c,d\in\mathbb{C};\hspace{.15cm}\ABS{c}\leq a;\hspace{.15cm}\ABS{d}\leq b .
\end{equation}
For the case $a>b$, which we encounter here, it can be shown that the concurrence defined in Eq.~\eqref{eq:Concurrence} reduces to
\begin{equation}
\label{eq:result_Con}
C = \begin{cases}
2\left(\ABS{c}-b\right) ,& \ABS{c}>b \\
0 ,& \ABS{c}\leq b .
\end{cases}
\end{equation}

The behavior of the concurrence at $T=4$\,K shown in Fig.~\ref{fig:phase_transition}(a) is thus directly linked to the difference between two elements in the two-photon density matrix. With increasing driving strength, the elements corresponding to $c$ are reduced, while the occupations $b$ are enhanced. Thus the concurrence decreases with rising driving strength and drops to zero once the condition $\ABS{c}=b$ is reached, i.e., when the coherence corresponding to $\OP{HH}{VV}$ is reduced to the level of the occupations of $\KET{HV}$ and $\KET{VH}$. Afterwards, the concurrence remains zero as $c$ decreases further. This indicates the existence of a critical driving strength, where the generated photons transition from an entangled photon pair to a nonentangled one.
Note, that in the case without constant laser excitation the situation $\ABS{c}\leq b$ can never occur. Starting from an initially prepared biexciton, the states $\KET{HV}$ and $\KET{VH}$ cannot be reached in the undriven dynamics, i.e., $b=0$ in this case. Further note that when examining the two-photon density matrix at $\Omega_2$ [cf., Fig.~\ref{fig:phase_transition}(b) lower row, middle panel], one realizes that at $T=4\,$K the transition towards nonentangled photon emission occurs before the special point of the phonon-free curve, i.e., before the regime of $\Psi$BS is reached.

At the critical driving strength, a transition between entangled and nonentangled photon emission takes place, where the concurrence behaves similar to the order parameter in a phase transition: $C$ is finite below a certain driving strength and strictly zero above it. Because a vanishing concurrence has a well defined physical meaning, the character of the two-photon state, as described by the density matrix $\rho^\text{2p}$, changes qualitatively at the transition point. Due to the one-to-one correspondence between the concurrence and the entanglement of formation, a vanishing concurrence is linked to a distinct physical property of the generated two-photon state.

If the concurrence vanishes, the density matrix $\rho^\text{2p}$ can be decomposed into a statistical mixture
\begin{equation}
\rho^\text{2p} = \sum\limits_j p_j \PRO{\psi_j} ,
\end{equation}
where all pure two-photon states $\KET{\psi_j}=\KET{\psi_{j,1}}\otimes\KET{\psi_{j,2}}$ can be factorized into a product of single-photon states which describe only the first and second detected photon, respectively. $p_j$ is the probability to find the system in the corresponding pure state. For the two-photon density matrix of the form in Eq.~\eqref{eq:special_form}, an explicit expression for a possible decomposition is provided in Appendix~\ref{app:decomposition}.
Therefore, above the critical driving strength, the two-photon state can always be expressed as a statistical mixture of factorizeable states. This property changes when the driving is below its critical value, where no such decomposition is possible. The practical implication of a system being in a factorizable state is that performing a measurement on the first photon has no implication on the outcome of a subsequent measurement on the second photon, in sharp contrast to what is found for an entangled state.

Note that the transition point, i.e., the critical driving strengths, depends on the chosen delay time window $\tau$. In general, the critical driving strength increases with shorter $\tau$ values. However, changing the delay time window between 1 and 100~ps did not result in a qualitatively different behavior of the concurrence, and in all cases the concurrence at $T=4$~K dropped to zero before the special point at $\Omega=\Omega_2$.

\subsection{Phonon influence on the two-photon density matrix}
\label{subsec:phonon_influence}

Because the transition between entangled and nonentangled photon emission as well as the behavior of the concurrence can be traced back to the form of the two-photon density matrix, the remaining task is to understand the phonon influence on the system and on $\rho^\text{2p}$. To this end, we consider the laser-dressed states $\chi\in \lbrace U,M,N,L\rbrace$ accompanied by two-photon states with different combinations for the photon polarizations, i.e., system states $\KET{\chi,2,0}$, $\KET{\chi,1,1}$ and $\KET{\chi,0,2}$. Here a state $\KET{\chi,n\st{H},n\st{V}}$ denotes a state with $n\st{H}$ ($n\st{V}$) horizontally (vertically) polarized photons inside the cavity. The numerical results for the driving strength $\Omega=\Omega_3$ show that two-photon states associated with a fixed laser-dressed state enter the steady state of the system with almost equal occupation probabilities but vanishing coherences between them. This steady state results directly in the obtained two-photon density matrix that can be well described by
\begin{equation}
\begin{split}
\label{eq:2pdm_num}
\rho^\text{2p} = \frac{1}{6}
\begin{pmatrix}
2 & 0 & 0 & 0 \\
0 & 1 & 1 & 0 \\
0 & 1 & 1 & 0 \\
0 & 0 & 0 & 2 
\end{pmatrix}
\end{split} .
\end{equation}

In the following we provide a simplified argument why the interaction with LA phonons leads to this type of steady state and two-photon density matrix. In the absence of further relaxation processes, e.g., cavity losses and radiative decay, phonons lead to a steady state which is close to a thermal distribution over the eigenstates $\KET{\varphi_\nu}$ of the system Hamiltonian without the phonon contribution  that can be reached from the initial state \cite{PI_phonon-assisted,PI_undressing,GlaessL2012_relaxation}. Here, the Hamiltonian describing the driven QD-cavity system without phonons is $\hat{H}_0:=\hat{H}\st{QD-C}+\hat{H}\st{L}$. The analyses in Ref.~\onlinecite{Seidelmann_QUTE_2020} suggest that the interaction with the cavity modes introduces a weak coupling between laser-dressed states with different numbers of photons inside the cavity. Hence the eigenstates of the full Hamiltonian $\hat{H}_0$ are best described by
\begin{equation}
\KET{\varphi_\nu} = \sum\limits_{\chi,n\st{H},n\st{V}} a_\nu(\chi, n\st{H}, n\st{V}) \KET{\chi,n\st{H},n\st{V}} ,
\end{equation}
where the mixing coefficients $a_\nu(\chi, n\st{H}, n\st{V})$ depend on the cavity coupling strength $g$, and the energetic placement of the cavity modes, i.e., the cavity-laser detuning $\Delta$.

The most important two-photon states in a thermalized distribution, which should define the character of the two-photon density matrix $\rho^\text{2p}$, are the energetically lowest ones. In the considered system, these are the three two-photon states accompanied by the laser-dressed state $\KET{L}$, i.e., $\KET{L,2,0}$, $\KET{L,1,1}$, and $\KET{L,0,2}$. These three states are precisely the ones distinct by the chosen (two-photon) resonance condition. Using a Schrieffer-Wolff transformation \cite{Winkler,LossDivincenzo_SW}, an effective Hamiltonian describing this two-photon resonance can be constructed. The explicit epression for this effective Hamiltonian, in the basis $\KET{U,0,0}$, $\KET{L,1,1}$, $\KET{L,\Phi_+}=(\KET{L,2,0}+\KET{L,0,2})/\sqrt{2}$, and $\KET{L,\Phi_-}=(\KET{L,2,0}-\KET{L,0,2})/\sqrt{2}$, is \cite{Seidelmann_QUTE_2020}
\begin{equation}
\label{app:eq:result_2p_UL_v3}
\hat{\tilde{H}}\st{UL}^{(2)} 
\approx E\st{U} \mathds{1}_4
+ g ^2  \begin{pmatrix}
\delta^{\text{UL}} & \gamma_1^{\text{UL}} & -\gamma_2^{\text{UL}} & 0\\
\gamma_1^{\text{UL}} & -\delta_2^{\text{UL}} & \alpha^{\text{UL}} & 0\\
-\gamma_2^{\text{UL}} & \alpha^{\text{UL}} & -\delta_2^{\text{UL}} & 0\\
0 & 0 & 0 & -\delta_2^{\text{UL}}
\end{pmatrix}
\end{equation}
with
\begin{eqnarray}
 c &=&\dfrac{2\Omega}{\sqrt{8\Omega^2+ \left( \Delta_0 + \sqrt{\Delta_0^2+8\Omega^2} \right)^2}} \notag \\ 
\tilde{c} &=& \sqrt{\frac{1}{2}-c^2} \notag \\
\delta^{\text{UL}} &=& \left(\tilde{c}^2-c^2\right) \left(\frac{2}{\Delta_0}+\frac{4 \left(\tilde{c}^2-c^2\right)}{\dcL{UL}}\right) \notag \\
\delta_2^{\text{UL}} &=& \delta^{\text{UL}}+\delta_3^\text{UL} \notag \\
\delta_3^\text{UL} &=& \frac{8\left(\tilde{c}^2-c^2\right)^2}{3\Delta_\text{UL}} + \frac{2\tilde{c}^2}{\Delta_\text{UL}+\Delta_0/2} +\frac{2c^2}{\Delta_\text{UL}-\Delta_0/2} \notag\\
\gamma_1^{\text{UL}} &=& 4c \tilde{c}\frac{1}{\Delta_0}  - 16 c\tilde{c}\left(\tilde{c}^2- c^2\right)\frac{1}{\dcL{UL}} \notag \\ 
\gamma_2^{\text{UL}} &=& 16 c\tilde{c}\left(\tilde{c}^2- c^2\right)\frac{1}{\dcL{UL}} \notag \\
\alpha^{\text{UL}} &=& \frac{1}{\Delta_0} - \left(1 -16 c^2 \tilde{c}^2\right)\frac{1}{\dcL{UL}} -\frac{1}{2}\delta_3^\text{UL} +\frac{2\tilde{c}^2}{\Delta_\text{UL}+\Delta_0/2}  \,, \notag 
\end{eqnarray}
where $\mathds{1}_4$ is the four-dimensional identity matrix. All couplings associated with two-photon processes ($\gamma$'s and $\alpha$'s) and energy shifts ($\delta$'s) in this effective Hamiltonian are on the order of $g^2/\Delta_0$. Therefore, the energetic splitting between the four eigenstates  $\KET{\varphi_\nu^\text{UL}}$ of $\hat{\tilde{H}}\st{UL}^{(2)}$ is on the same order. For our realistic parameters, the energy $g^2/\Delta_0 = 0.1g \approx 5$~$\mathrm{\mu eV}$ is already two orders of magnitude smaller than the thermal energy $k_B T$ at $T=4$~K. Consequently, in a thermalized distribution all four eigenstates $\KET{\varphi_\nu^\text{UL}}$ should appear with the same weights. 

Note that, if no further loss mechanisms are considered, the state $\KET{L,\Phi_-}$ is decoupled from the inital state $\KET{G,0,0}$ and cannot occur in a thermalized distribution \cite{GlaessL2012_relaxation}. But, due to cavity losses and radiative decay, this state can be reached during the system dynamics and, thus, should appear in our situation. In general, the exact steady state of systems with both type of relaxation mechanisms, LA phonon-induced relaxation and further loss processes, is difficult to predict and may differ qualitatively from what is expected in the limiting cases where only one type of relexation mechanism is considered \cite{PI_nonHamil2016}.

However, under the assumption that phonon-induced relaxations dominate over the latter, a steady state should emerge, where the leading two-photon contribution is proportional to a thermal distribution over all four eigenstates $\KET{\varphi_\nu^\text{UL}}$:
\begin{equation}
\label{eq:rho_therm_UL}
\begin{split}
\hat{\rho}\st{th}^\text{UL} =& \frac{1}{Z}\sum\limits_{\nu=1}^4 \exp\left[-\frac{\epsilon_\nu^\text{UL}}{k_B T}\right] \PRO{\varphi_\nu^\text{UL}} \\
\approx& \frac{1}{4} \sum\limits_{\nu=1}^4 \PRO{\varphi_\nu^\text{UL}} \\
=& \frac{1}{4} ( \PRO{U,0,0} + \PRO{L,1,1} \\
&+ \PRO{L,2,0} + \PRO{L,0,2} ) ,
\end{split}
\end{equation}
where $\epsilon_\nu^\text{UL}$ are the corresponding energies.

One can calculate a first approximation for the two-photon density matrix due to this contribution
\begin{equation}
\begin{split}
\label{eq:2pdm_therm_UL}
(\rho\st{th}^\text{2p})_{jk,\ell m} &\approx \mathcal{N} \Tr{\hat{a}_m\hat{a}_\ell \hat{\rho}\st{th}^\text{UL} \hat{a}_j^\dagger\hat{a}_k^\dagger} = \frac{1}{6}
\begin{pmatrix}
2 & 0 & 0 & 0 \\
0 & 1 & 1 & 0 \\
0 & 1 & 1 & 0 \\
0 & 0 & 0 & 2 
\end{pmatrix} ,
\end{split}
\end{equation}
where $\mathcal{N}$ is a normalization constant and one assumes that the delay time window is small, justifying the limit $\tau\rightarrow0$ for this estimate. 

This thermal two-photon density matrix $\rho\st{th}^\text{2p}$ is also depicted in Fig.~\ref{fig:phase_transition}(c) for comparison. This density matrix is almost identical to the one obtained for $\Omega_3$ at $T=4$\,K. Thus, although the presented argument is only a simplified analysis, it fits quite well to the numerical observations: (i) two-photon states with different combinations of the polarization appear with the same occupation probabilities and vanishing coherences, c.f., Eq.~\eqref{eq:rho_therm_UL}. (ii) The two-photon density matrix $\rho\st{th}^\text{2p}$ associated with a phonon-induced thermalization is close to the one encountered in the numerical simulations.

This implies that the role of phonons is a thermalization in laser-driven eigenstates. Furthermore, the two-photon density matrix at $\Omega_2$ is also close to $\rho\st{th}^\text{2p}$, the main difference being small, finite coherences. Thus, in this case, the thermalization process is not complete. This points to a competition between different relaxation processes that, each on their own, lead to different steady states in the system dynamics.
While the interaction with LA phonons on its own leads to a thermalization and, consequently, to the thermal two-photon density matrix $\rho\st{th}^\text{2p}$, the other relaxation processes, i.e., cavity losses and radiative decay, result in the steady state associated with the phonon-free result given in the row above.
The imprint of this competition is even more prominent in the two-photon density matrix for the driving strength $\Omega_1$ at $4$\,K. Here, $\rho^\text{2p}$ describes a two-photon state roughly in between the two limiting cases, that are given by the corresponding phonon-free result and $\rho\st{th}^\text{2p}$.

Consequently, the two-photon state at $T=4$\,K is a result of two competing relaxation mechanisms, that, each on their own, would result in a different steady state in the system dynamics, but that act on different time scales. While the time scale of the cavity losses and radiative decay is independent of the driving strength $\Omega$, the time scale of the thermalization is reduced with increasing $\Omega$, and thus the two-photon density matrix approaches its thermal limit. The reason for this reduction lies in the phonon spectral density. As the driving strength increases, the transition energies between the laser-dressed states increase as well. Therefore, the main contribution of the phonon spectral density $J(\omega)$ has to be evaluated at a higher frequency. In the parameter range considered in this work, this results in a higher value of $J$ which translates to a stronger average phonon coupling, cf. Appendix~\ref{app:gaas_density}. Consequently the thermalization process should become more dominant with increasing driving strength, pulling the two-photon density matrix closer to $\rho\st{th}^\text{2p}$.

\begin{figure}[t]
\centering
\includegraphics[width=\columnwidth]{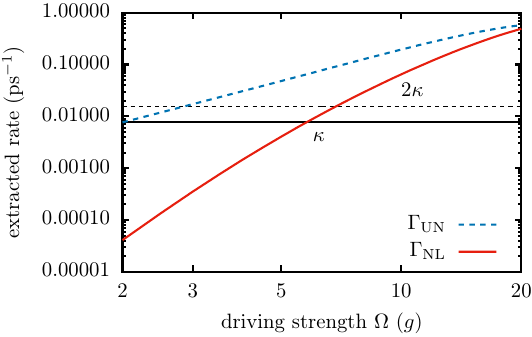}
\caption{
Phonon-induced transition rate $\Gamma\st{UL}$ ($\Gamma\st{NL}$) between the laser-dressed states $\KET{U}$ and $\KET{L}$ ($\KET{N}$ and $\KET{L}$) as a function of the driving strength $\Omega$. Additionally, a dashed (solid) horizontal black line indicates the value of the cavity loss rate $2\kappa$ ($\kappa$) associated with a two-photon (one-photon) state.
}
\label{fig:rates}
\end{figure}

The rates leading to thermalization are estimated using Fermi's golden rule for the phonon-induced rates $\Gamma_{\chi\chi^\prime}$ between the laser-dressed states $\KET{\chi}$ and $\KET{\chi^\prime}$ with $\chi,\chi^\prime\in\lbrace U,M,N,L\rbrace$ associated with phonon emission and absorption processes. Based on this estimate, two finite rates $\Gamma\st{UN}$ and $\Gamma\st{NL}$ are extracted, cf. Appendix~\ref{app:phonon_rates}, which  are shown in Fig.~\ref{fig:rates} as a function of the driving strength.  
As expected, in the considered parameter range, both rates increase with $\Omega$, supporting the previous argument on the increasing role of phonons. Interestingly, shortly after $\Omega\approx 5g$, i.e., the driving strength where the concurrence vanishes, both rates  become larger than the cavity loss rate $\kappa$. Above this point both phonon-related rates exceed cavity or radiate loss rates in the system. Thus, according to this estimate, the phonon-induced thermalization dominates the system dynamics for higher driving strength values, leading to a vanishing entanglement.

\subsection{Markovian rate model}
\label{subsec:PD_model}

So far, all results were obtained within the numerical complete path-integral method. However, it is worth exploring whether similar results can also be obtained when a Markovian rate model is used to account for the influence of LA phonons. Based on the assumption that the main impact of phonons is indeed captured by the phonon-induced rates between the laser-dressed states extracted in Fig.~\ref{fig:rates}, we replace the exact phonon Hamiltonian $\hat{H}\st{Ph}$ in Eq.~\eqref{eq:LvN_Eq} with a Lindblad-type operator
\begin{equation}
\label{eq:rate_model}
\begin{split}
\mathcal{L}\st{Ph}\hat{\rho} =& \LB{\OP{N}{U}}{(n\st{UN}+1)\Gamma\st{UN}}\hat{\rho} + \LB{\OP{U}{N}}{n\st{UN}\Gamma\st{UN}}\hat{\rho}\\
&+ \LB{\OP{L}{N}}{(n\st{NL}+1)\Gamma\st{NL}}\hat{\rho} + \LB{\OP{N}{L}}{n\st{NL}\Gamma\st{NL}}\hat{\rho}
\end{split}
\end{equation}
that describes the relevant phonon-induced transitions. Here
\begin{equation}
n_{\chi\chi^\prime} = \left( \exp\left[\frac{E_\chi-E_{\chi^\prime}}{k\st{B}T}\right] -1 \right)^{-1}
\end{equation}
is the expected number of phonons with the required energy $E_\chi-E_{\chi^\prime}$ excited at temperature $T$ according to the Bose-Einstein statistics. Note that, besides the rates $\Gamma_{\chi\chi^\prime}$, also the dressed-state energies and compositions depend on the driving strength $\Omega$.

Although this rate model does not match the results of the path-integral simulations exactly, the qualitative behavior of the concurrence is reproduced, cf., black crosses in Fig.~\ref{fig:phase_transition}(a). This finding strengthens the interpretation that the main origin of the observed transition is indeed a competition between the phonon-induced relaxation that leads to a thermalization and the other relaxation processes (i.e., cavity losses and radiative decay). However, for driving strength values between $3g$ and $5g$, the rate model underestimates the phonon influence. This is to be expected because the Markovian rate model in Eq.~\eqref{eq:rate_model} does not capture non-Markovian effects related to the pure dephasing-type coupling to LA phonons, that are typically important in strongly-confined QDs.

\subsection{Temperature dependence at fixed driving strengths}
\label{subsec:crit_temp}

\begin{figure*}[t!]
\centering
\includegraphics[width=\textwidth]{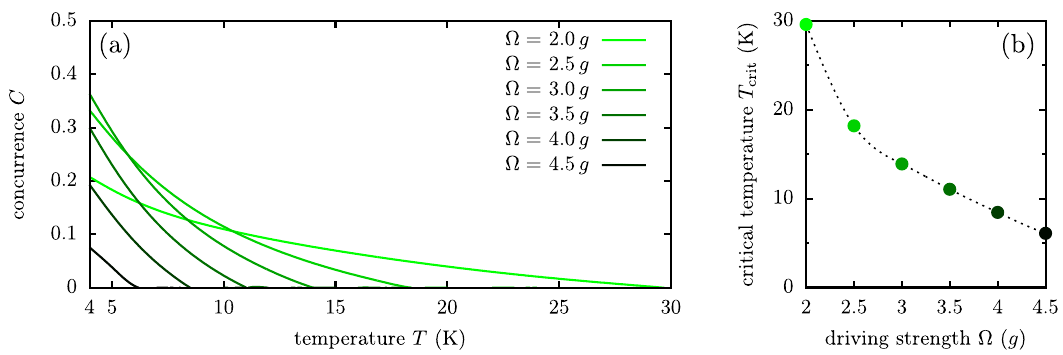}
\caption{
(a) Concurrence as a function of the temperature (above $4\,$K) for different driving strength values $\Omega$ between  $2.0g$ and $4.5g$. (b) Critical temperature $T_\text{crit}$ of the phonon-induced phase transition for the driving strength values in panel (a). The dashed line is a guide-to-the-eye.
}
\label{fig:crit_temp}
\end{figure*}

In this section, the temperature dependence at fixed driving strength values is investigated. As the average number of phonons that is excited at a given temperature increases with rising $T$, an increasing phonon impact is to be expected in this situation. Thus, similar to the critical driving strength in Fig.~\ref{fig:phase_transition}(a), also a critical temperature should exist (for a given $\Omega$), where the transition between entangled and nonentangled photon emission takes place.

In Fig.~\ref{fig:crit_temp}(a), the concurrence is shown as a function of the temperature $T$ for different driving strength values between $2.0g$ and $4.5g$. The results indeed demonstrate that a phonon-induced transition is taking place. At a given driving strength, the concurrence decreases monotonically with rising temperature before vanishing at a certain critical temperature. The concurrence remains zero when the temperature is further increased. Furthermore, as the driving strength increases, the drop becomes more rapid. Note that we restrict our numerical calculations to temperatures above 4~K, i.e., the temperature of liquid helium, because the path-integral calculations become more demanding at lower temperatures and might not be fully converged. However, calculations at 2~K suggest that also at this temperature the concurrence remains zero in the $\Psi$BS regime for $\Omega>6.12g$.

Figure~\ref{fig:crit_temp}(b) depicts the critical temperatures calculated for the six driving strength values in panel (a). It is evident that the critical temperature decreases with rising driving strength $\Omega$. As shown in the previous sections, a higher driving strength leads to larger phonon-induced rates and faster thermalization. Thus the critical transition temperature is the lower the higher the applied driving strength. The results imply that even moderate degrees of entanglement in the constantly driven system can be achieved only at temperatures below 30\,K. Furthermore at temperatures of 4\,K and above the generation of a $\Psi$BS seems to be no longer possible. We find that, for realistic parameters, phonon interactions lead to a vanishing concurrence for driving strengths far below the value needed to switch from $\Phi$BS to $\Psi$BS. Therefore, we observe no $\Psi$BS in simulations accounting for phonons for all parameters considered here.

The average number of phonons in the thermal equilibrium and, thus, the phonon influence increases with rising temperature. As a result, the competition between loss processes, i.e., cavity losses and radiative decay, on one hand and the thermalization due to phonons on the other hand, is more and more dominated by the latter. Therefore the degree of entanglement decreases with rising temperature as the two-photon density matrix is again pulled towards the thermal one. At a certain, critical temperature $T\st{crit}$ the transition takes place and the degree of entanglement, i.e., the concurrence, drops to zero. At this critical temperature, the coherence between $\KET{HH}$ and $\KET{VV}$ is reduced to the level of the occupations $\KET{HV}$ and $\KET{VH}$, i.e., the two-photon density matrix can be expressed as a statistical sum of factorizable two-photon states. Because the phonon influence only increases at larger temperatures, the concurrence remains zero.

\begin{figure}[t]
\centering
\includegraphics[width=\columnwidth]{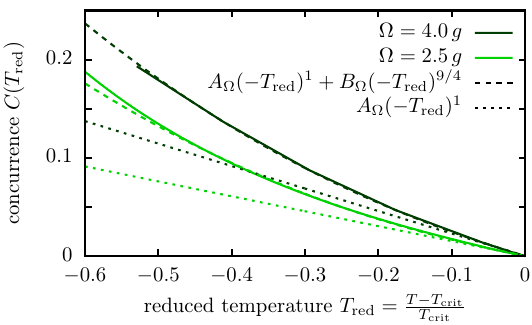}
\caption{
Concurrence as a function of the reduced temperature for two different driving strength values $\Omega = 4.0g$ and $2.5g$ (dark and light green solid lines, respectively). In addition, corresponding fits to power laws containing only the exponent $k=1$ (doted lines) and one additional term (dashed lines) are shown with the same color. 
}
\label{fig:crit_exp}
\end{figure}

Because the observed transition between entangled and nonentangled photon emission, displays features typically associated with a phase transition, we also investigate the concurrence as a function of the reduced temperature $T\st{red}=(T-T\st{crit})/T\st{crit}$.
In the case of an actual phase transition, the behavior of the order parameter when approaching the critical temperature should be given by a power law described by a critical exponent, independent of other system parameters like the driving strength. Quite remarkably, we indeed find that the concurrence obeys a power law  
\begin{equation}
\label{eq:def_crit_exp}
C \propto \left( \frac{T}{T\st{crit}}-1 \right)^k ,
\end{equation}
where the exponent $k=1$ is the same for all driving strength values considered in Fig.~\ref{fig:crit_temp}. Thus, also in this aspect, the concurrence resembles the order parameter of a phase transition. To illustrate this, Fig.~\ref{fig:crit_exp} depicts the concurrence as a function of the reduced temperature for two selected driving strength values. 
In both cases, when the temperature approaches the corresponding critical value, the concurrence is described by a power law with the exponent $k=1$ and fitted amplitudes $A_\Omega$.

We also observe an interesting side aspect: when the temperature is decreased further, the concurrence evaluated for the two driving strengths still exhibits the same behavior. Both curves can be well described by a power law extended by an additional term with a fitted exponent 9/4 and amplitudes $B_\Omega$. This feature holds for all driving strength values considered in Fig.~\ref{fig:crit_temp} (not shown).

Note that, although the discussed transition resembles a phase transition in some aspects, other characteristics typically associated with a phase transition could not be observed. In particular, neither a discontinuity in some other quantity besides the derivative of the concurrence nor a critical slowing down could be found.

\section{Conclusion} 
\label{sec:conclusion}

In conclusion, we have investigated the influence of phonons on entangled photon pairs generated from a constantly driven QD-cavity system. We find a strong reduction and a severe qualitative impact on the degree of entanglement, as measured by the concurrence, already at temperatures as low as $4\,$K. Unlike in the phonon-free situation, the influence of phonons suppresses the generation of $\Psi$BS entanglement in the studied parameter range, even at low temperatures. The concurrence decreases with increasing temperature and driving strength until a critical parameter value is reached where it drops to zero and remains so at larger values.

Note that, although we have limited the discussion to the two-photon resonance condition between the uppermost and lowest laser-dressed state where the highest degrees of entanglement had been obtained in the phonon-free situation, we checked that the qualitative behavior of the concurrence at other possible two-photon resonances is the same: at a certain driving strength or temperature the degree of entanglement drops to zero and remains so afterward. Furthermore, also choosing the negative sign in Eq.~\eqref{eq:fix_Delta} does not alter the results qualitatively. 

The observed transition between entangled and nonentangled photon emission is akin to a phase transition where the concurrence resembles the order parameter. We encounter a phonon-induced transition, which cannot take place in a situation without constant laser excitation. The reason behind this phenomenon is a competition between two different mechanisms, a thermalization due to phonons and other loss processes, i.e., radiative decay and cavity losses, that determines the steady state of the system. A higher temperature or driving strength gives rise to a stronger phonon impact. Eventually, the thermalization dominates, driving the two-photon state towards a thermal state with vanishing degree of entanglement. Thus the emitted photon pair transforms from an entangled towards a nonentangled state.

Because the transition can be traced back to a thermalization due to phonon-induced transitions, entanglement at higher driving strengths and/or temperature, in particular a $\Psi$BS, might be observable in situations where the pure dephasing-type coupling to LA phonons is not so effective. This is the case in larger, weakly-confined QDs, since the effective phonon coupling, as described by the phonon spectral density, decreases with rising QD dimensions.
However, to decide whether or not this decrease is enough to make the $\Psi$BS observable,  requires an extended investigation that accounts for possible influences of higher energy states, since the energetic spacing between adjacent QD states is smaller in the weak-confinement limit. Apart from this, even in strongly-confined QDs the $\Psi$BS might reappear for much larger driving strength values (which cannot be investigated within current implementations of the path-integral formalism due to time-discretization issues) when the energy differences between the laser-dressed states exceeds the range of the phonon spectral density, resulting in a decoupling of the LA phonons.

\section*{Acknowledgments}
D. E. Reiter acknowledges support by the Deutsche Forschungsgemeinschaft (DFG) via the project 428026575.
We are further greatful for support by the Deutsche Forschungsgemeinschaft (DFG, German Research Foundation) via the project 419036043.

\appendix

\section{GaAs parameters and phonon spectral density}
\label{app:gaas_density}

\begin{figure}[t]
\centering
\includegraphics[width=\columnwidth]{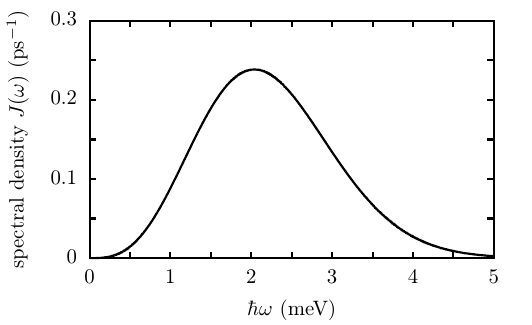}
\caption{
Phonon spectral density for the considered GaAs QD with electron (hole) confinement length $a\st{e}=3$~nm ($a\st{h} = a\st{e}/1.15$) and material parameters listed in Table~\ref{app:tab:GaAs_Parameters}.
}
\label{app:fig:J}
\end{figure}

In this paper, a spherically symmetric self-assembled GaAs QD with a harmonic oscillator confinement and an electron (hole) confinement length $a\st{e}=3$~nm ($a\st{h} = a\st{e}/1.15$) is considered. Furthermore a linear dispersion relation is assumed for the LA phonons. In this situation, the explicit expression for the phonon spectral density $J(\omega)$ is \cite{PI_undressing,PI_nonHamil2016}
\begin{equation}
\label{app:eq:J}
J(\omega) = \frac{\omega^3}{4\pi^2\hbar\rho\st{D}c\st{S}^5} \left[ D\st{e} \,e^{-\omega^2 a\st{e}^2/(4c\st{S}^2)} - D\st{h} \,e^{-\omega^2 a\st{h}^2/(4c\st{S}^2)} \right]^2 .
\end{equation}
The necessary material parameters are taken from literature \cite{Krummheuer2005} and listed in Table~\ref{app:tab:GaAs_Parameters}. For these parameters, the resulting spectral density is shown in Fig.~\ref{app:fig:J}. Note that using a spherical QD model provides for the calculation of the reduced density matrix as considered here no loss of generality as shown in Ref.~\onlinecite{Lueker2017_Dot_geometry}.

\begin{table}
\centering
\caption{GaAs material parameters taken from Ref.~\onlinecite{Krummheuer2005}.}
\label{app:tab:GaAs_Parameters}
\begin{ruledtabular}
\begin{tabular}{l c c}
Parameter & & Value\\
\hline
Mass density ($\mathrm{kg/m^3}$) & $\rho\st{D}$ & 5370 \\
Sound velocity ($\mathrm{m/s}$) & $c\st{S}$ & 5110 \\
Electron deformation potential (eV) & $D\st{e}$ & 7.0 \\
Hole deformation potential (eV) & $D\st{h}$ & -3.5
\end{tabular}
\end{ruledtabular}
\end{table}

\section{Decomposition in factorizable states}
\label{app:decomposition}

Due to the interaction with LA phonons, we encounter two-photon density matrices that can, in good approximation, be described as 
\begin{equation}
\label{app:eq:special_form}
\rho^\text{2p} = \begin{pmatrix}
a & 0 & 0 & c\\
0 & b & d & 0\\
0 & d^\ast & b & 0\\
c^\ast & 0 & 0 & a
\end{pmatrix} ,
\end{equation}
in the basis $\lbrace \KET{H_1H_2}, \KET{H_1V_2}, \KET{V_1H_2}, \KET{V_1V_2}\rbrace$, where the index refers to the first or second detected photon.
The parameters fulfill the requirements for an arbitrary density matrix
\begin{equation}
a,b\in\mathbb{R}^+_0;\hspace{.15cm}2(a+b)=1;\hspace{.15cm}c,d\in\mathbb{C};\hspace{.15cm}\ABS{c}\leq a;\hspace{.15cm}\ABS{d}\leq b .
\end{equation}
In the case $a>b$, the corresponding concurrence is given by
\begin{equation}
\label{app:eq:result_Con}
C = \begin{cases}
2\left(\ABS{c}-b\right) ,& \ABS{c}>b \\
0 ,& \ABS{c}\leq b .
\end{cases}
\end{equation}
Because the concurrence has a one-to-one correspondence to the entanglement of formation, obtaining a vanishing concurrence has a well-defined physical meaning: in this situation, there exists at least one decomposition of the density matrix
\begin{equation}
\rho^\text{2p} = \sum\limits_j p_j \PRO{\psi_j} ,
\end{equation}
where all pure (two-photon) states $\KET{\psi_j}$ factorize into quantum states that describe only the first or second detected photon. $p_j$ is the probability to encounter this pure state in the mixed state described by $\rho^\text{2p}$. Since the concurrence vanishes for $\ABS{c}\leq b$, such a decomposition must exist in this situation. Here, we give an explicit expression for a possible decomposition.

After introducing the phase $\varphi$ and $\theta$ for the parameter $c=\ABS{c}e^{i\varphi}$ and $d=\ABS{d}e^{i\theta}$, respectively, we re-write the two-photon density matrix. In the situation $a\geq b \geq \ABS{c}$, our decomposition depends on the relation between $\ABS{c}$ and $\ABS{d}$.

In the case $\ABS{d}\geq\ABS{c}$, we obtain the following possible decomposition
\begin{equation}
\label{app:eq:expansion_dc}
\begin{split}
\rho^\text{2p} =& 2\left( a - \ABS{d} \right) \rho^\text{2p}_1 + 2\left( b-\ABS{d} \right) \rho^\text{2p}_2 + 4\ABS{c} \rho^\text{2p}_3 \\
&+ 2\left( \ABS{d}-\ABS{c} \right)\left( \rho^\text{2p}_4 + \rho^\text{2p}_5 \right) ,
\end{split}
\end{equation}
where all contributions $\rho^\text{2p}_j$ to the density matrix can be expressed as a mixed state
\begin{equation}
\rho^\text{2p}_j = \frac{1}{2} \left( \PRO{\psi_j^{(\alpha)}} + \PRO{\psi_j^{(\beta)}} \right)
\end{equation}
containing two factorizable pure states $\KET{\psi_j^{(\alpha/\beta)}}$. The explicit expressions for these quantities are
\begin{subequations}
\begin{equation}
\rho^\text{2p}_1 = \frac{1}{2}
\begin{pmatrix}
1 & 0 & 0 & 0 \\
0 & 0 & 0 & 0 \\
0 & 0 & 0 & 0 \\
0 & 0 & 0 & 1
\end{pmatrix}
\end{equation}
\begin{equation}
\KET{\psi_1^{(\alpha)}} = \KET{H_1}\KET{H_2};\hspace{.15cm}\KET{\psi_1^{(\beta)}} = \KET{V_1}\KET{V_2}
\end{equation}
\begin{equation}
\rho^\text{2p}_2 = \frac{1}{2}
\begin{pmatrix}
0 & 0 & 0 & 0 \\
0 & 1 & 0 & 0 \\
0 & 0 & 1 & 0 \\
0 & 0 & 0 & 0
\end{pmatrix}
\end{equation}
\begin{equation}
\KET{\psi_2^{(\alpha)}} = \KET{H_1}\KET{V_2};\hspace{.15cm}\KET{\psi_2^{(\beta)}} = \KET{V_1}\KET{H_2}
\end{equation}
\begin{equation}
\rho^\text{2p}_3 = \frac{1}{4}
\begin{pmatrix}
1 & 0 & 0 & e^{i\varphi} \\
0 & 1 & e^{i\theta} & 0 \\
0 & e^{-i\theta} & 1 & 0 \\
e^{-i\varphi} & 0 & 0 & 1
\end{pmatrix}
\end{equation}
\begin{equation}
\label{app:eq:psi3ab}
\begin{split}
\KET{\psi_3^{(\alpha/\beta)}} =& \frac{1}{\sqrt{2}} \left( \KET{H_1} \pm e^{-i(\varphi+\theta)/2}\KET{V_1} \right) \\
&\times \frac{1}{\sqrt{2}} \left( \KET{H_2} \pm e^{-i(\varphi-\theta)/2}\KET{V_2} \right)
\end{split}
\end{equation}
\begin{equation}
\rho^\text{2p}_{4/5} = \frac{1}{4}
\begin{pmatrix}
1 & 0 & 0 & \pm 1 \\
0 & 1 & e^{i\theta} & 0 \\
0 & e^{-i\theta} & 1 & 0 \\
\pm 1 & 0 & 0 & 1
\end{pmatrix} .
\end{equation}
\end{subequations}
The contribution $\rho^\text{2p}_4$ ($\rho^\text{2p}_5$), exhibiting the positive (negative) coherence $\OP{H_1 H_2}{V_1 V_2}$, is a special case of $\rho^\text{2p}_3$ with $\varphi=0$ ($\varphi=\pi$). Thus the corresponding pure states $\KET{\psi_{4/5}^{(\alpha/\beta)}}$ are given as special cases of Eq.~\eqref{app:eq:psi3ab}.

In the case $\ABS{d}<\ABS{c}$, a possible decomposition can be constructed in a slightly different form
\begin{equation}
\label{app:eq:expansion_cd}
\begin{split}
\rho^\text{2p} =& 2\left( a - \ABS{c} \right) \rho^\text{2p}_1 + 2\left( b-\ABS{c} \right) \rho^\text{2p}_2 + 4\ABS{d} \rho^\text{2p}_3 \\
&+ 2\left( \ABS{c}-\ABS{d} \right)\left( \rho^\text{2p}_6 + \rho^\text{2p}_7 \right) .
\end{split}
\end{equation}
Again the contributions
\begin{equation}
\rho^\text{2p}_{6/7} = \frac{1}{4}
\begin{pmatrix}
1 & 0 & 0 & e^{i\varphi} \\
0 & 1 & \pm 1 & 0 \\
0 & \pm 1 & 1 & 0 \\
e^{-i\varphi} & 0 & 0 & 1
\end{pmatrix}
\end{equation}
are special cases of $\rho^\text{2p}_3$ with $\theta=0$ and $\theta=\pi$, respectively. Thus they, can be decomposed into a sum over two factorizable pure states $\KET{\psi_{6/7}^{(\alpha/\beta)}}$ which are special cases of $\KET{\psi_{3}^{(\alpha/\beta)}}$.

Altogether, in the situation $a\geq b \geq \ABS{c}$, the two-photon density matrix in Eq.~\eqref{app:eq:special_form} can be always expressed as a sum over 10 or less factorizable pure (two-photon) states 
\begin{equation}
\label{app:eq:general_decmp}
\rho^\text{2p} = \sum\limits_j p_j \frac{1}{2} \left( \PRO{\psi_j^{(\alpha)}} + \PRO{\psi_j^{(\beta)}} \right)
\end{equation}
where the probabilities $p_j$ are the corresponding (real and positive) prefactors in the expansion \eqref{app:eq:expansion_dc} or \eqref{app:eq:expansion_cd}, respectively. Because at least one decomposition into factorizable pure states exists, the corresponding entanglement of formation is zero and, in turn, the concurrence vanishes.

\section{Phonon-induced transition rates between laser-dressed states}
\label{app:phonon_rates}

In this section, we estimate the phonon-induced rates that lead to a thermalization of the system. To this end, we consider the four laser-dressed states \cite{Seidelmann_QUTE_2020}
\begin{subequations}
\label{app:eq:dressed states}
\begin{equation}
\KET{U} = c \left( \KET{G}+\KET{B} \right) + \tilde{c} \left( \KET{X\st{H}}+\KET{X\st{V}} \right)
\end{equation}
\begin{equation}
\KET{M} = \frac{1}{\sqrt{2}} \left( \KET{X\st{H}}-\KET{X\st{V}} \right)
\end{equation}
\begin{equation}
\KET{N} = \frac{1}{\sqrt{2}} \left( \KET{G}-\KET{B} \right)
\end{equation}
\begin{equation}
\KET{L} = \tilde{c} \left( \KET{G}+\KET{B} \right) - c \left( \KET{X\st{H}}+\KET{X\st{V}} \right)
\end{equation}
\begin{equation}
\label{app:eq:coeffs}
 c = \dfrac{2\Omega}{\sqrt{8\Omega^2+ \left( \Delta_0 + \sqrt{\Delta_0^2+8\Omega^2} \right)^2}}; \hspace{0.2cm} \tilde{c} = \sqrt{\frac{1}{2}-c^2} ,
\end{equation}
\end{subequations}
where $\Omega$ is the driving strength and $\Delta_0$ the energetic detuning between exciton states and laser. The corresponding energies $E_\chi$ are given in Eq.~\eqref{eq:laser_dressed energies}.

According to Fermi's golden rule, the phonon-induced rates $\Gamma_{i\rightarrow f}$ between an initial laser-dressed state with zero phonons $\KET{i}$ and a final dressed state at lower energy with one phonon $\KET{f}$ can be estimated as
\begin{equation}
\label{app:eq:FGR}
\begin{split}
\Gamma_{i\rightarrow f} = \frac{2\pi}{\hbar} \ABS{\BRA{f} \hat{H}\st{Ph} \KET{i}}^2 g(E_f) ,
\end{split}
\end{equation}
where $g(E_f)$ is the density of states at the energy of the final state. In our situation, using the phonon spectral density, Eq.~\eqref{app:eq:FGR} can be reformulated as
\begin{equation}
\Gamma_{\chi\chi^\prime} = 2\pi \,\ABS{\BRA{\chi} \hat{V}\st{Ph} \KET{\chi^\prime}}^2 \,J( \ABS{E_\chi-E_{\chi^\prime}}/\hbar) ,
\end{equation}
where $\chi,\chi^\prime \in\lbrace U,M,N,L\rbrace$ and the operator $\hat{V}\st{Ph}$ takes the form
\begin{equation}
\hat{V}\st{Ph} =
\begin{pmatrix}
1 & 0 & \sqrt{2}c & 0 \\
0 & 1 & 0 & 0 \\
\sqrt{2}c & 0 & 1 & \sqrt{2}\tilde{c} \\
0 & 0 & \sqrt{2}\tilde{c} & 1
\end{pmatrix}
\end{equation}
in the basis $\lbrace \KET{U}, \KET{M}, \KET{N}, \KET{L}\rbrace$.

Using this estimate, we obtain two finite rates
\begin{subequations}
\begin{equation}
\Gamma\st{UN} = 4\pi c^2 J([E\st{U}-E\st{N}]/\hbar)
\end{equation}
\begin{equation}
\Gamma\st{NL} = 4\pi \tilde{c}^2 J([E\st{N}-E\st{L}]/\hbar)
\end{equation}
\end{subequations}
that are associated with transitions between laser-dressed states $\KET{U}\leftrightarrow\KET{N}$ and  $\KET{N}\leftrightarrow\KET{L}$ due to phonon  emission or absorption processes. 
These rates depend on the dressed-state energies and the coefficients $c,\tilde{c}$ given by Eq.~\eqref{app:eq:coeffs} and  thus, in particular, on the driving strength $\Omega$.

\bibliography{PIbib}

\end{document}